# Long-Distance Spin Transport Through a Graphene Quantum Hall Antiferromagnet

Petr Stepanov[1,2*], Shi Che[1,2*], Dmitry Shcherbakov[1,2], Jiawei Yang[1,2], Kevin Thilahar[1], Greyson Voigt[1], Marc W. Bockrath[1,2], Dmitry Smirnov[3], Kenji Watanabe[4], Takashi Taniguchi[4], Roger K. Lake[5†], Yafis Barlas[1,5‡], Allan H. MacDonald[6§], Chun Ning Lau[1,2**]

[1] Department of Physics and Astronomy, University of California, Riverside, CA 92521
[2] Department of Physics, The Ohio State University, Columbus, OH 43221
[3] National High Magnetic Field Laboratory, Tallahassee, FL 32310
[4] National Institute for Materials Science, 1-1 Namiki Tsukuba Ibaraki 305-0044 Japan.
[5] Department of Electrical and Computer Engineering, University of California, Riverside, CA 92521
[6] Department of Physics, University of Texas at Austin, Austin, TX 78712-1192
* These authors contribute equally to this work.

**Antiferromagnetic insulators (AFMI) are robust against stray fields, and their intrinsic dynamics could enable ultrafast magneto-optics and ultrascaled magnetic information processing. Low dissipation, long distance spin transport and electrical manipulation of antiferromagnetic order are much sought-after goals of spintronics research. Here, we report the first experimental evidence of robust long-distance spin transport through an AFMI, in our case the gate-controlled, canted antiferromagnetic (CAF) state that appears at the charge neutrality point of graphene in the presence of an external magnetic field. Utilizing gate-controlled quantum Hall (QH) edge states as spin-dependent injectors and detectors, we observe large, non-local electrical signals across a 5 µm-long, insulating channel only when it is biased into the $\nu$=0 CAF state. Among possible transport mechanisms, spin superfluidity in an antiferromagnetic state gives the most consistent interpretation of the non-local signal's dependence on magnetic field, temperature and filling factors. This work also demonstrates that graphene in the QH regime is a powerful model system for fundamental studies of ferromagnetic and antiferromagnetic spintronics.**

An important goal of spintronics research is to identify mechanisms that minimize dissipation in devices that seek to exploit the action of spin currents. In magnetic insulators, spin-currents can be carried dissipatively by magnon quasiparticles[1-3]. In the case of systems with easy plane magnetic order, they can also be carried collectively in the form of dissipationless spin supercurrents[4-9]. While magnon transport is much less efficient in an ideal antiferromagnetic insulators (AFMI) than that in ferromagnetic insulators in the absence of a thermal gradient, spin superfluidity is theoretically expected to be a possibility in both cases. Although the potential of AFM materials[5,6,10] as active spintronic components that can be electrically manipulated has been recognized[11], and important progress has been made in demonstrating theoretically expected properties[10,12-16], spin transport through an AFMI thicker than ~10 nm has yet to be demonstrated.

---

[†] Email: rlake@ece.ucr.edu
[‡] Email: yafisb@ucr.com
[§] Email: macd@physics.utexas.edu
[**] Email: lau.232@osu.edu

In a parallel thread of scientific progress, monolayer graphene has emerged as a versatile platform to investigate quantum Hall (QH) physics. In strong magnetic fields, the approximate *SU(4)* spin-valley invariance symmetry is frequently broken, giving rise to gate tunable order[17-38]. For example, whereas states at Landau level filling factor $\nu=\pm2$ do not have broken symmetries and support two co-propagating chiral edge channels with opposite spins, states at $\nu=\pm1$ are spontaneously spin-polarized and support a single spin-polarized chiral edge channel. The $\nu=0$ state of neutral graphene is particularly interesting because it has no counterpart in the traditional GaAs-based quantum Hall systems, and is a true insulator with no edge states and longitudinal *and* transverse resistances that diverge in the low-temperature limit. The consensus emerging from experiment is that the $\nu=0$ state is a canted AFM, with nearly opposite easy-plane spin polarizations on graphene's two sub-lattices[24,34]. However, a direct demonstration of the AFM order has been missing prior to the present work.

Because of its gate-tunable magnetic order and its extremely weak spin-orbit coupling, high quality graphene in the QH regime has been proposed as an attractive model system for fundamental spintronics studies[9,39]. In this paper, we implement the proposal made in ref. [9], using the $\nu=0$ state of graphene as the AFM insulator and combining the $\nu=-2$ and $-1$ states to realize spin injectors, and detectors. We detect a large non-local voltage signal, up to 300 μV, that is transmitted ~ 5 μm across the AFM insulator. The signal disappears when filter regions are tuned away from $\nu=-1$ where they supply spin-dependence. Both the magnitude of the non-local signal and the transport distances are orders of magnitudes larger than in oxide-based AFM insulators[10,12-16], suggesting that a fundamentally different mechanism is at play.

The basic operating principle and geometry of the device is illustrated in Fig. 1a-c[9]. It consists of a graphene sheet in the *x-y* plane with a series of separately contacted top gates that is placed in a large perpendicular magnetic field *B*. The central top gate $V_{cTg}$ is used to tune the graphene region underneath to $\nu=0$, at which the ground state is an AFM insulator[17-37]. Zeeman coupling to the AFM order leads to A and B sublattice spins that are nearly in the x-y plane, but not quite oppositely oriented because of slight canting toward the field direction. The canting angle $\theta$ depends on the ratio of the Zeeman energy to the valley-flip isospin anisotropy energy, as explained below. To the left and right of the CAF state are the injection and detector regions, respectively, each consisting of a top-gated region flanked by two "bare" (non-top-gated) regions. During device operation, the filling factors of the left (right) top-gated regions are tuned to $\nu_{inj}=-1$ ($\nu_{det}=-1$) at which the ground state is a ferromagnetic QH insulator with a conducting chiral edge channel that is fully spin-polarized opposite to the net magnetization direction, and the bare regions are tuned to $\nu=-2$ which has a non-magnetic QH ground state that supports chiral edge channels of both spins. In the injection region, a voltage bias $V_{bias}$ is applied between the 2 "bare" regions so as to establish a chemical potential difference between the transmitted ↑ and reflected ↓ spin channels. When impinging upon the CAF region, the incident spin current at the left CAF interface can produce a spin transfer torque that favors the formation of spiral (Néel) spin textures that carry spin current collectively[4,7,9,40]. In the detector region, this spin current is converted into a spin-polarized charge current via a reciprocal process and measured as a non-local voltage $V_{nl}$ between the two voltage probes attached to the two $\nu=-2$ regions.

Because the magnetic states of graphene in the QH regime are susceptible to disorder[18], realization of this proposal requires fabrication of devices of exceptional quality[37]. With this goal, we have assembled a long monolayer encapsulated in hexagonal BN sheets (Fig. 1c). The heterostructure is constructed using a pick-up technique and coupled to Cr/Au electrodes via 1D edge contacts[41]. Three independent top gates are deposited on the device. The length $L$ and with $W$ of the central top gate, the CAF region during device operation, are ~ 5 μm and 2.5 μm, respectively. The charge densities of the 4 "bare" regions are controlled by the back-gate voltage $V_{Bg}$, while those of the top-gated regions are tuned by both back and top gates. Since each gate can be independently modulated, up to 4 different carrier densities can be created within the device. Moreover, every region is attached to 1-2 pairs of electrical leads, to enable its independent electrical characterization. In our best device, the delicate ferromagnetic QH state at $\nu$=-1 is resolved at $B$>6T, and the base temperature longitudinal resistivity of the insulating $\nu$=0 state increases from 1 MΩ to 2.3 MΩ (Fig. 1d) when the magnetic field strength is increased from 13T to 18T.

Our main experimental findings are presented in Figs. 2a-c. We first explore spin transport by independently modulating the filling factors $\nu_{inj}$ and $\nu_{det}$ of the top-gated injector and detector regions, while conserving $\nu$=0 and $\nu$=-2 states at the central top-gated and the bare regions, respectively. We apply a bias voltage $V_{bias}$=0.4V while monitoring the non-local signal $V_{nl}$ (see device configuration in Fig. 2d). Fig. 2a plots $V_{nl}$ (color) at $B$=18T as $\nu_{inj}$ and $\nu_{det}$ vary from -3 to +1. Prominent signals are observed *only* when both $\nu_{inj}$ and $\nu_{det}$ are tuned to be close to $\nu$=-1, *i.e.* only when the detecting and injecting regions contain spin-filters. The dark blue area for $\nu_{det}$>-0.5 signifies amplifier saturation when the right top-gated region enters the insulating AFM state. Figs. 2b-c show the individual line traces of $V_{nl}$ for fixed $\nu_{inj}$=-1 and fixed $\nu_{det}$=-1, respectively. The non-local signal detected across the 5 μm channel is exceedingly large, with a maximum amplitude of ~225 μV centered at $\nu_{inj}$=$\nu_{det}$=-1.

A number of different physical mechanisms could be responsible for the non-local signal: charge tunneling, percolation, drift or diffusion, a spin Seebeck effect carried by thermal magnons, or spin superfluidity in the easy-plane AFMI $\nu$=0 channel. In the remaining portion of the manuscript, we present data that either supports or undermines these different scenarios. We find that the spin-superfluidity mechanism is the one that is consistent with all of the data.

If the non-local signal was due to charge transport, either via tunneling, percolation, drift or diffusion, one would expect that that $V_{nl}$ would increase as the channel conductivity increases, since the charge could then more easily traverse the channel. When the magnetic field is reduced from 18T to 15T, the conductivity of the $\nu$=0 channel *increases* by a factor of 1.4 (Fig. 1c). However, $V_{nl}$ decreases by a factor of 2.8 to ~80 μV (yellow line trace, Fig. 2b). The *decrease* in $V_{nl}$ as the $\nu$=0 state becomes more *conductive* contradicts the trend expected from a charge leakage mechanism.

To further confirm that the non-local signal arises from transport of a pure spin current through the AFM, we perform a control study in which the central region is tuned instead to the insulating $\nu$=+2 state, which like the $\nu$=-2 state has unpolarized chiral edge channels (Fig. 2f). The $V_{nl}(\nu_{inj}, \nu_{det})$ map at $B$=18T (Fig. 2e) for this case indicates minimal response. Thus, the non-local signal is small when any one of the 3 regions (the injector, the detector, or the center region) is tuned to a $\nu$=-2 or $\nu$=+2 state. Taken together, these control measurements demonstrate that non-local signals indeed arise from spin transport, and not drift, diffusion, or percolation of charge currents.

We note that non-local signals at the Dirac point in graphene have been observed previously[42], but differ dramatically in origin from those in the current work. In ref. [42], the entire device is gated into the $\nu=0$ state, and the non-local signals, which persisted at high temperature and low magnetic fields, were attributed to long-range flavor Hall effects, though they could also arise from magneto-thermoelectric effects[43]. In contrast, our devices are specifically configured with the $\nu=-2/-1/-2$ regions for spin injection and detection, and the non-local signal appears only in the low-temperature, high-magnetic-field regime where the AFMI state forms. Moreover, our control measurements rule out the Zeeman spin and valley Hall effects.

We have also examined the dependence of the non-local signal on the bias voltage $V_{bias}$ that controls the electro-chemical potential difference between the up and down spins in the injector. Here we apply the same top gate voltage to both injector and detector top gates, so that $\nu_{inj}=\nu_{det}$ throughout the measurements, while maintaining the "bare" regions at $\nu=-2$ and the central top-gated region at $\nu=0$. Fig. 3a presents $V_{nl}$ as a function of $V_{bias}$ and the filling factors of the injector and detector regions. As before, prominent signals are observed for $\nu_{inj}=\nu_{det}=-1$. For $V_{bias}>0$, the signal is approximately linear in $V_{bias}$ (Fig. 3b). The smallest value of $V_{bias}$ at which non-local signal is first observed is $<\sim0.01$ V, and is limited by resolution of the sweep. This small $V_{bias}$ value, together with the linear dependence, confirms that Joule heating is not a concern. On the other hand, since the non-local voltage is expected to be linear in $V_{bias}$ for both spin superfluid and charge current mechanisms[9], this observation does not on its own rule out either mechanism.

Finally, we compare the temperature dependence of the non-local signal with that of the QH effects. At $B=18$T, the magnitude of the non-local signal decreases with increasing $T$, and disappears at ~ 45 K. Since spin transport in our devices depends critically on the QH states in graphene, we also measure $R_{xx}(V_{Bg})$ at $B=18$T and temperatures ranging from 2.7 K to 85K. As illustrated in Fig. 3c, the $\nu=0$ and $\nu=-2$ states are very robust and remain well quantized even at $T=85$K. The $\nu=-1$ state is more fragile; its $R_{xx}$ increases with $T$ and the QH effect is barely resolved when $T$ increases to ~35 K. To compare the dependences of the QH states and the non-local signal, in Fig. 3d we plot $R_{xx}$ (right axis, blue triangles) and $V_{nl}$ (left axis, red squares) versus temperature in Arrhenius scales as a function of $1/T$. Both data sets can be satisfactorily fit to a thermal activation model, with characteristic temperatures of 23.8±2.2K and 24.8±2.4K, respectively. The similar temperature dependences strongly suggest that the non-local transport signals in the range of 2 to 40K are dependent on the spin-filtering action of the $\nu=-1$ state, that enables both spin injection and detection. At the same time, the temperature dependence of $V_{nl}$ is opposite to that expected from a spin Seebeck effect mediated by thermal magnons.

We now further consider the two spin transport mechanisms mediated by magnetic order. Spins can be carried either by magnon quasiparticles or collectively in the form of spin supercurrents. In an *ideal* AFM, magnons do not carry spin. However, the magnetic order of the $\nu=0$ QH state is slightly canted in the z-direction, so the z-injected spin current could drive magnons that diffuse across the AFMI, transporting spins from one end to the other. The spin carried by a magnon is proportional to the canting angle θ, which is very small. According to a previous study of the CAF state in graphene[34], the ratio of the Zeeman energy to the valley-flip anisotropy energy is ~25, suggesting that $θ<3°$. Thus, canting-induced magnon transport of spin across the 5 μm AFMI is insignificant. Also, as noted above, the monotonic decline of the non-local signals with increasing temperature is opposite to that expected from magnon-mediated transport, as magnons are thermally activated and should be more effective at higher temperatures.

Because of the above considerations, and the spin transport distance that is $10^3$-$10^4$ times longer than previous studies of magnons in oxide-based AFMIs[10,12-16], we conclude that a fundamentally different mechanism underlies our experimental results. Thus, we consider the only other known spin transport mechanism in an AFMI -- coherent Néel textures that allow superfluid transport of spins polarized in the *z*-direction (see Auxiliary Supplementary Materials for an animation of such spin transport). In the limit of large spin stiffness, efficient spin-injection, and weak violation of valley-projected number conservation, the non-local voltage is predicted[44] to satisfy

$$\frac{V_{nl}}{V_{bias}} = \frac{F_{inj} g_{inj}}{F_{inj} g_{inj} + \alpha \frac{e^2}{h} \frac{A}{2\pi l_B^2}} \quad (1)$$

as the easy-plane ferromagnet case considered in Ref.[44] and the easy-plane antiferromagnet case relevant here have identical spin-superfluid responses to injected spin-currents. Here $g_{inj}$ is the longitudinal conductance of the $\nu$=-1 regions that is expected to be very close to $e^2/h$, $l_B$ is the magnetic length, $F_{inj}$ is the efficiency factor for spin-injection, $A$ is the area of the CAF region, and $\alpha$ is the magnetization damping parameter. To estimate $F_{inj}$, it is important to consider how charge is carried along the perimeter of the $\nu$=-2 regions, which are surrounded on three sides by $\nu$=0 regions and therefore support two hole-like chiral edge channels. For boundaries between $\nu$=-2 and vacuum these channels have spins polarized along and opposite to the magnetic field. However, for the critical boundary between the $\nu$=-2 region and the $\nu$=0 CAF, the edge channels in the narrow boundary limit will be those of the occupied quasiparticle states from the broken symmetry N=0 Landau level which have spin-polarization close to the *x-y* plane. Because these quasiparticles cannot carry z-polarized spins, $F_{inj}$ can be close to 1 if the boundary can be made sharp. Since, in our experiments $\frac{V_{nl}}{V_{bias}}$~$10^{-4}$ and the CAF area $A$=12.5 μm$^2$, we obtain from Eq. (1) that $\alpha$ ~$10^{-2}$ $F_{inj}$, consistent with $\alpha$ values in the $10^{-4}$ to $10^{-2}$ range typical of AFMs[45-47]. Magnetization damping in the $\nu$=0 AFM state is likely due to decay channels opened up by density inhomogenities within the sample[48,49].

The applied voltage, gate voltage, magnetic field, and temperature dependencies of the non-local voltage are all consistent with the mechanism of superfluid spin transport across a 5-μm AFMI state. Other possible interpretations of our observations are contradicted by one or more trends of the data. In particular, the *decrease* of $V_{nl}$ by a factor of 2.8 when the magnetic field is decreased from 18T to 15T while the $\nu$=0 state conductivity *increases* by a factor of 1.4 rules out a simple charge leakage mechanism; the decrease of $V_{nl}$ with increasing temperature rules out a spin Seebeck mechanism mediated by thermal magnons. Lastly, the manifestation of non-local signal only when the injector and detector are made spin-selective by introducing $\nu$=-1 regions and when the channel is tuned to the $\nu$=0 AFM insulating state rules out the Zeeman spin and valley Hall effects. The negative signal at $\nu_{det}$=-3 (Fig. 2 a-b) is particularly intriguing, as it suggests a different spin texture in high LLs[31]. Taken together, we therefore attribute the observed robust non-local signal across the 5-μm AFMI state in graphene to collective spin transport.

In summary, using the $\nu$=-1 spin polarized edge states as injectors and detectors, we demonstrated long-distance spin transport through the $\nu$=0 insulating state in graphene, with all of the experimental evidence consistent with spin superfluid transport through an AFMI.

Detection of a large, non-local signal over a distance of 5 μm is particularly exciting. The results also suggest that graphene as a remarkably tunable model system for investigating antiferromagnetic spintronics. Many open questions, such as the length, width and mobility dependence of the spin transport signals, the efficiency of the spin-injection mechanism and its dependence on gating geometry, the mechanisms for spin scattering and spin loss, the AFMI state in bilayer graphene with the additional layer degree of freedom, await experimental and theoretical investigation.

**References**

1   Tsoi, M. *et al.* Generation and detection of phase-coherent current-driven magnons in magnetic multilayers. *Nature* **406**, 46-48 (2000).
2   Neusser, S. & Grundler, D. Magnonics: Spin Waves on the Nanoscale. *Adv. Mater.* **21**, 2927-2932 (2009).
3   Chumak, A. V., Vasyuchka, V. I., Serga, A. A. & Hillebrands, B. Magnon spintronics. *Nat. Phys.* **11**, 453-461 (2015).
4   Takei, S., Moriyama, T., Ono, T. & Tserkovnyak, Y. Antiferromagnet-mediated spin transfer between a metal and a ferromagnet. *Phys. Rev. B* **92**, 020409 (2015).
5   Jungwirth, T., Marti, X., Wadley, P. & Wunderlich, J. Antiferromagnetic spintronics. *Nat. Nanotechnol.* **11**, 231-241 (2016).
6   Baltz, V. *et al.* Antiferromagnetism: the next flagship magnetic order for spintronics ? *preprint*, arXiv:1606.04284 (2016).
7   Takei, S., Halperin, B. I., Yacoby, A. & Tserkovnyak, Y. Superfluid spin transport through antiferromagnetic insulators. *Phys. Rev. B* **90**, 094408 (2014).
8   Konig, J., Bonsager, M. C. & MacDonald, A. H. Dissipationless spin transport in thin film ferromagnets. *Phys. Rev. Lett.* **87**, 187202 (2001).
9   Takei, S., Yacoby, A., Halperin, B. I. & Tserkovnyak, Y. Spin Superfluidity in the nu=0 Quantum Hall State of Graphene. *Phys. Rev. Lett.* **116**, 216801 (2016).
10  Wadley, P. *et al.* Electrical switching of an antiferromagnet. *Science* **351**, 587 (2016).
11  Núñez, A. S., Duine, R. A., Haney, P. & MacDonald, A. H. Theory of spin torques and giant magnetoresistance in antiferromagnetic metals. *Phys. Rev. B* **73**, 214426 (2006).
12  Wang, H. L., Du, C. H., Hammel, P. C. & Yang, F. Y. Antiferromagnonic Spin Transport from Y3Fe5O12 into NiO. *Phys. Rev. Lett.* **113**, 097202 (2014).
13  Hahn, C. *et al.* Conduction of spin currents through insulating antiferromagnetic oxides. *EPL* **108**, 57005 (2014).
14  Moriyama, T. *et al.* Anti-damping spin transfer torque through epitaxial nickel oxide. *Appl. Phys. Lett.* **106**, 162406 (2015).
15  Wang, H. L., Du, C. H., Hammel, P. C. & Yang, F. Y. Spin transport in antiferromagnetic insulators mediated by magnetic correlations. *Phys. Rev. B* **91**, 220410 (2015).
16  Lin, W., Chen, K., Zhang, S. & Chien, C. L. Enhancement of Thermally Injected Spin Current through an Antiferromagnetic Insulator. *Phys. Rev. Lett.* **116**, 186601 (2016).
17  Yang, K., Sarma, S. D. & MacDonald, A. H. Collective modes and skyrmion excitations in graphene SU(4) quantum Hall ferromagnets. *Phys. Rev. B* **74**, 075423 (2006).
18  Nomura, K. & MacDonald, A. H. Quantum Hall ferromagnetism in graphene. *Phys. Rev. Lett.* **96**, 256602 (2006).

19  Alicea, J. & Fisher, M. P. A. Graphene integer quantum Hall effect in the ferromagnetic and paramagnetic regimes. *Phys. Rev. B* **74**, 075422 (2006).
20  Abanin, D. A. *et al.* Dissipative quantum Hall effect in graphene near the Dirac point. *Phys. Rev. Lett.* **98**, 1968006 (2007).
21  Fertig, H. A. & Brey, L. Luttinger liquid at the edge of undoped graphene in a strong magnetic field. *Phys. Rev. Lett.* **97**, 116805 (2006).
22  Goerbig, M. O., Moessner, R. & Doucot, B. Electron interactions in graphene in a strong magnetic field. *Phys. Rev. B* **74**, 161407 (2006).
23  Ostrovsky, P. M., Gornyi, I. V. & Mirlin, A. D. Theory of anomalous quantum Hall effects in graphene. *Phys. Rev. B* **77**, 195430 (2008).
24  Kharitonov, M. Edge excitations of the canted antiferromagnetic phase of the nu=0 quantum Hall state in graphene: A simplified analysis. *Phys. Rev. B* **86**, 075450 (2012).
25  Kim, S., Lee, K. & Tutuc, E. Spin-Polarized to Valley-Polarized Transition in Graphene Bilayers at ν=0 in High Magnetic Fields. *Phys. Rev. Lett.*, 016803 (2009).
26  Jiang, Z., Zhang, Y., Stormer, H. L. & Kim, P. Quantum Hall states near the charge-neutral Dirac point in graphene. *Phys Rev Lett* **99**, 106802 (2007).
27  Checkelsky, J. G., Li, L. & Ong, N. P. Zero-energy state in graphene in a high magnetic field. *Phys. Rev. Lett.* **100**, 206801 (2008).
28  Checkelsky, J. G., Li, L. & Ong, N. P. Divergent resistance at the Dirac point in graphene: Evidence for a transition in a high magnetic field. *Phys. Rev. B* **79**, 115434 (2009).
29  Giesbers, A. J. M. *et al.* Gap opening in the zeroth Landau level of graphene. *Phys. Rev. B* **80**, 201403 (2009).
30  Zhao, Y., Cadden-Zimansky, P., Jiang, Z. & Kim, P. Symmetry Breaking in the Zero-Energy Landau Level in Bilayer Graphene. *Phys. Rev. Lett.* **104**, 066801 (2010).
31  Young, A. F. *et al.* Spin and valley quantum Hall ferromagnetism in graphene. *Nat. Phys.* **8**, 550-556 (2012).
32  Amet, F., Williams, J. R., Watanabe, K., Taniguchi, T. & Goldhaber-Gordon, D. Selective Equilibration of Spin-Polarized Quantum Hall Edge States in Graphene. *Phys. Rev. Lett.* **112**, 196601 (2014).
33  Zhang, Y. *et al.* Landau-level splitting in graphene in high magnetic fields. *Phys. Rev. Lett.* **96**, 136806 (2006).
34  Young, A. F. *et al.* Tunable symmetry breaking and helical edge transport in a graphene quantum spin Hall state. *Nature* **505**, 528 (2014).
35  Bolotin, K. I., Ghahari, F., Shulman, M. D., Stormer, H. L. & Kim, P. Observation of the fractional quantum Hall effect in graphene. *Nature* **462**, 196-199 (2009).
36  Du, X., Skachko, I., Duerr, F., Luican, A. & Andrei, E. Y. Fractional quantum Hall effect and insulating phase of Dirac electrons in graphene. *Nature* **462**, 192-195 (2009).
37  Wu, F., Sodemann, I., Araki, Y., MacDonald, A. H. & Jolicoeur, T. SO(5) symmetry in the quantum Hall effect in graphene. *Phys. Rev. B* **90**, 235432 (2014).
38  Sun, Q.-f. & Xie, X. C. Spin-polarized $\nu=0$ state of graphene: A spin superconductor. *Phys. Rev. B* **87**, 245427 (2013).
39  Abanin, D. A., Lee, P. A. & Levitov, L. S. Spin-filtered edge states and quantum Hall effect in graphene. *Phys. Rev. Lett.* **96**, 176803 (2006).
40  Takei, S. & Tserkovnyak, Y. Superfluid Spin Transport Through Easy-Plane Ferromagnetic Insulators. *Phys. Rev. Lett.* **112**, 227201 (2014).

41  Wang, L. *et al.* One-Dimensional Electrical Contact to a Two-Dimensional Material. *Science* **342**, 614-617 (2013).
42  Abanin, D. A. *et al.* Giant Nonlocality Near the Dirac Point in Graphene. *Science* **332**, 328-330 (2011).
43  Studer, M. & Folk, J. A. Origins of Nonlocality Near the Neutrality Point in Graphene. **112**, 116601 (2014).
44  Chen, H., Kent, A. D., MacDonald, A. H. & Sodemann, I. Nonlocal transport mediated by spin supercurrents. *Phys. Rev. B* **90**, 220401 (2014).
45  Zhao, Y. *et al.* Experimental Investigation of Temperature-Dependent Gilbert Damping in Permalloy Thin Films. *Scientific Reports* **6**, 22890 (2016).
46  Johansen, Ø. & Linder, J. Current driven spin–orbit torque oscillator: ferromagnetic and antiferromagnetic coupling. *Scientific Reports* **6**, 33845 (2016).
47  Kim, T. H., Grünberg, P., Han, S. H. & Cho, B. Ultrafast spin dynamics and switching via spin transfer torque in antiferromagnets with weak ferromagnetism. *Scientific Reports* **6**, 35077 (2016).
48  Yankowitz, M. *et al.* Emergence of superlattice Dirac points in graphene on hexagonal boron nitride. *Nat. Phys.* **8**, 382-386 (2012).
49  Ju, L. *et al.* Photoinduced doping in heterostructures of graphene and boron nitride. *Nat. Nanotechnol.* **9**, 348-352 (2014).

**Acknowledgement**
We thank Hua Chen for helpful discussions. The work is supported by SHINES, which is an Energy Frontier Research Center funded by DOE BES under Award #SC0012670. AHM acknowledges partial support by the Welch foundation under grant TBF1473. SC is supported by DOE BES under Award # ER 46940-DE-SC0010597 to study QHE in graphene. Part of this work was performed at NHMFL that is supported by NSF/DMR-0654118, the State of Florida, and DOE. Growth of hexagonal BN crystals was supported by the Elemental Strategy Initiative conducted by the MEXT, Japan and a Grant-in-Aid for Scientific Research on Innovative Areas "Science of Atomic Layers" from JSPS.

**Fig. 1.** Device geometry, operating principle and characterization. **a.** Schematics of spin transport through the $\nu=0$ CAF state in graphene. The black, red and green arrows indicate Néel vectors of the AFM, polarization of transported spin, and direction of spin transport, respectively. **b.** Side-view schematic of device geometry. **c.** Optical image of the device. Scale bar = 5 μm. **d.** $R_{xx}(V_{bg})$ at the charge neutrality point for $B$=13T, 15T and 18T, respectively.

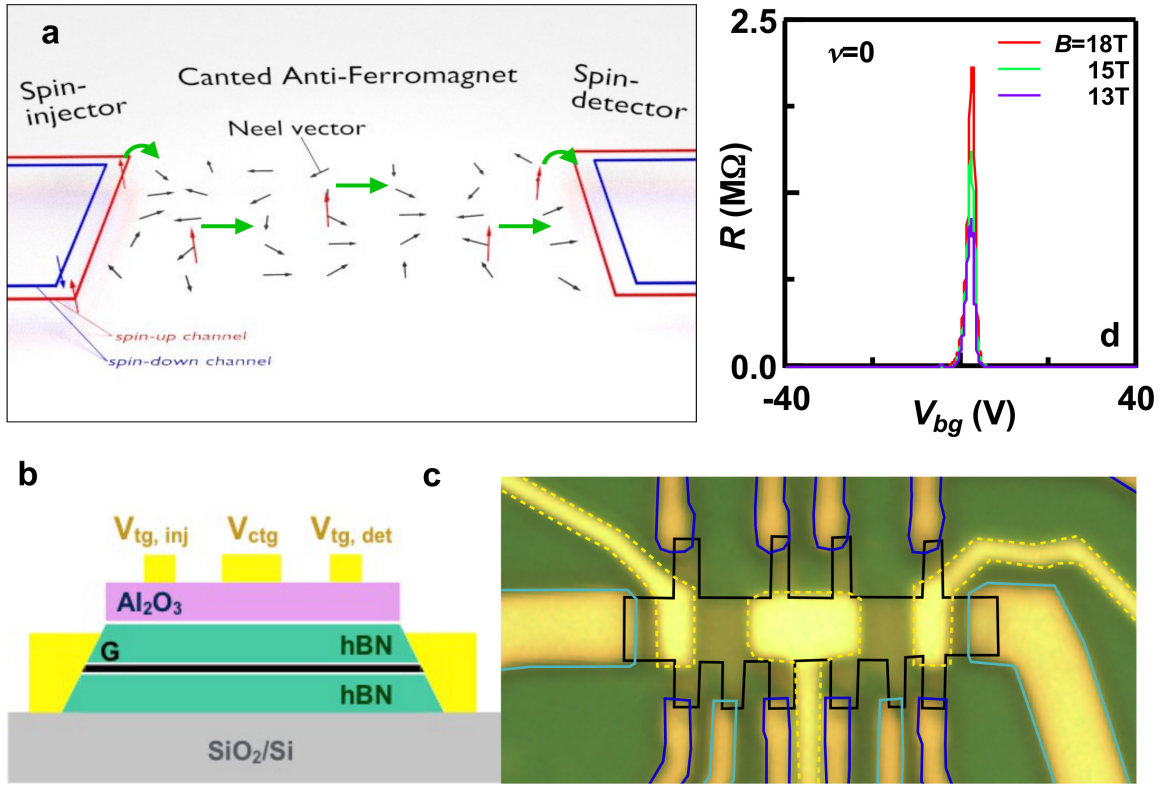

**Fig. 2.** Non-local transport data at $T$=260 mK. **a.** $V_{nl}(v_{inj}, v_{det})$ at $B$=18 T using the device configuration in (d), and line traces **b.** $V_{nl}(v_{det})$ at $v_{inj}$=-1 and **c.** $V_{nl}(v_{inj})$ at $v_{det}$=-1, respectively. The yellow curve in (b) is taken at $B$=15T. The line traces are offset by 30 µV to account for amplifier offset. **(e-f).** $V_{nl}(v_{inj}, v_{det})$ at $B$=18 T when the central top-gated region is set to $v$=2, using the device configuration in (f).

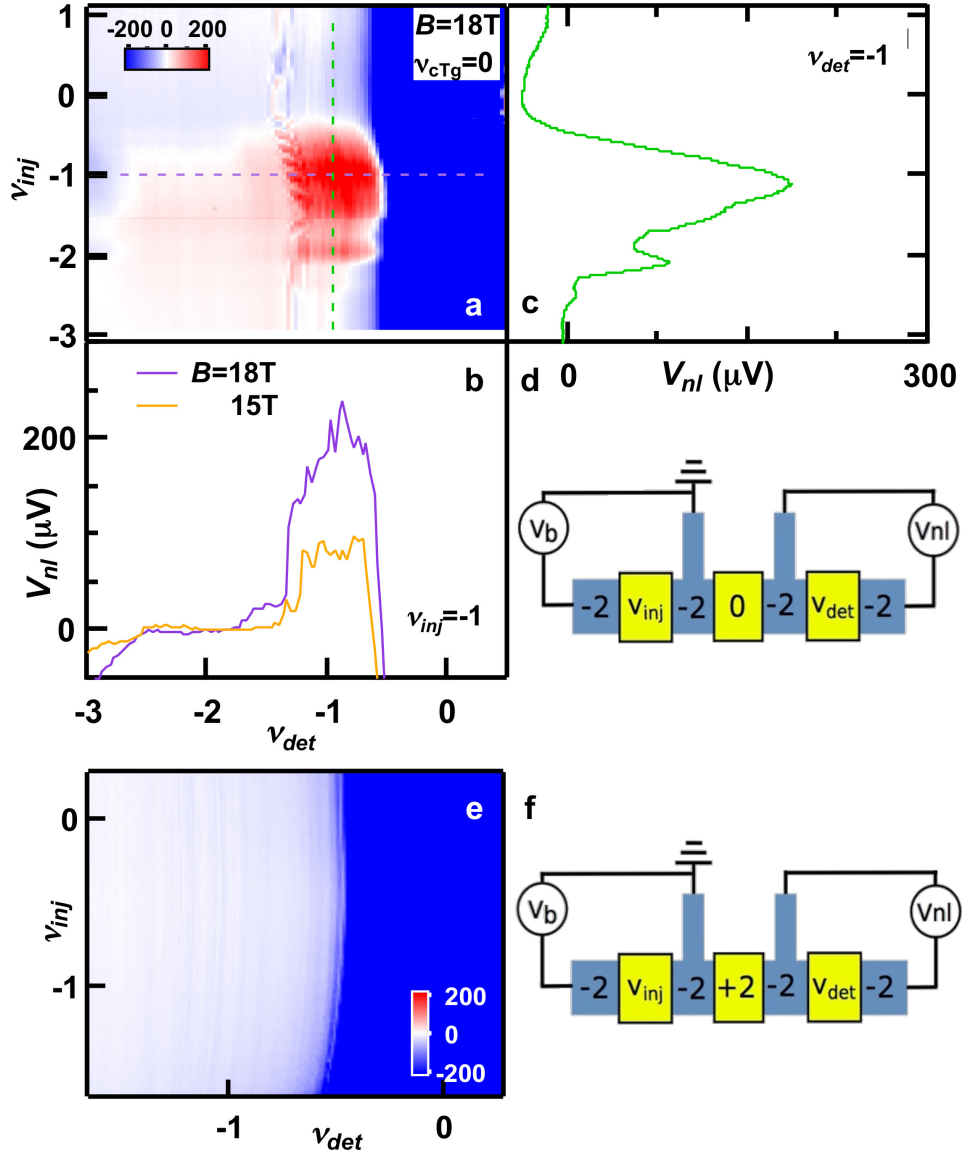

**Fig. 3.** Bias and temperature dependence of non-local signal. **a.** $V_{nl}(V_{bias}, \nu_{det})$ data at $B$=18T. **b.** A line cut $V_{nl}(V_{bias})$ along the dotted line. **c.** $R_{xx}(V_{Bg})$ at and $B$=18T and $T$=2.7, 5, 11, 14, 22, 27, 35, 45, 55, 70 and 84K (bottom to top). **d.** Maximum non-local signal at $\nu_{inj}=\nu_{det}$=-1 (left) and $R_{xx}(T)$ of the $\nu$=-1 state (right) plotted vs. $1/T$ on an Arrhenius scale. The solid lines are fits to thermal activation model.

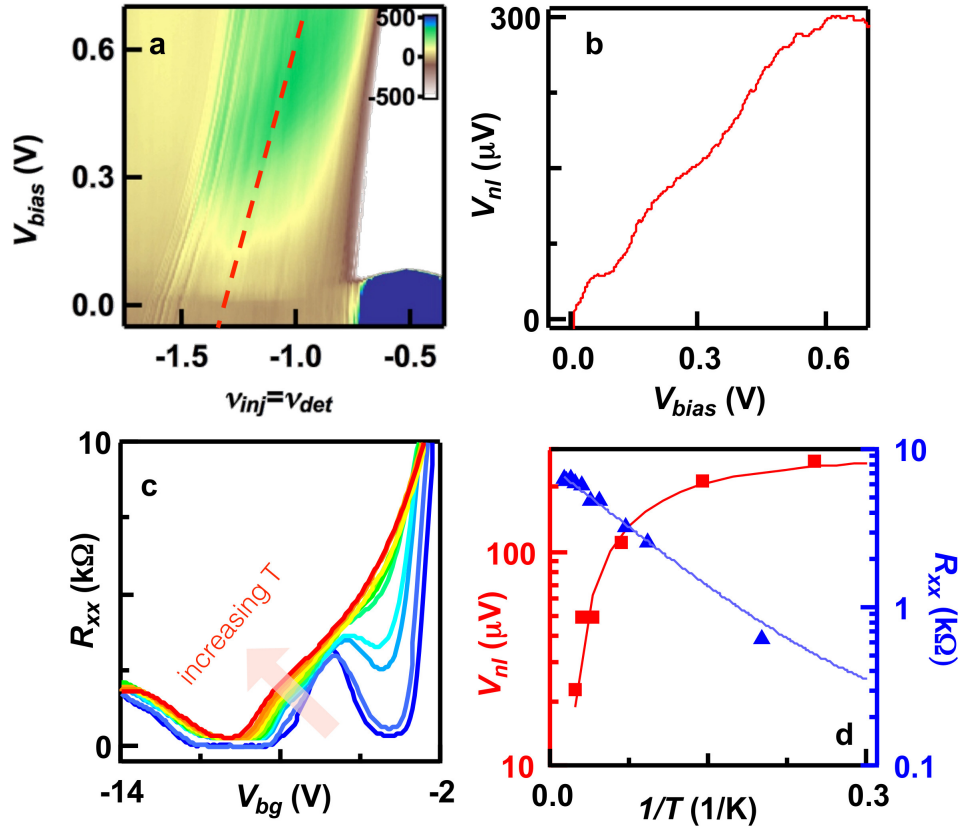